\documentstyle[twoside,12pt]{article}
\newcommand{\bp} {{\bf p}}

\newcommand{\btab}{\begin{tabbing}}
\newcommand{\etab}{\end{tabbing}}

\newcommand{\beqn}{\begin{equation}}
\newcommand{\eeqn}{\end{equation}}
\newcommand{\barr}[1]{\begin{array}{#1}}
\newcommand{\earr}{\end{array}}
\newcommand{\beqna}{\begin{eqnarray}}
\newcommand{\eeqna}{\end{eqnarray}}
\newcommand{\btablec}{\begin{table} \begin{center}}
\newcommand{\etablec}{\end{center} \end{table}}

\newcommand{\gapproxeq}{\lower.7ex\hbox{$\;\stackrel{\textstyle>}{\sim}\;$}}
\newcommand{\lapproxeq}{\lower.7ex\hbox{$\;\stackrel{\textstyle<}{\sim}\;$}} 
\newtheorem{theorem}{Theorem}
\newcommand{\bth}{\begin{theorem}}
\newcommand{\eth}{\end{theorem}}

\newcommand{\plabel}[1]{\label{#1}}
\newcommand{\pbibitem}[1]{\bibitem{#1}}
\newcommand{\cs}{{\cal S}}
\newcommand{\csbc}{{\cal S}_{B \leftrightarrow C}}
\newcommand{\cl}{{\cal L}}
\newcommand{\clbc}{{\cal L}_{B \leftrightarrow C}}

\newcommand{\cat}{{\cal A}_{tot}}
\newcommand{\cf}{{\cal F}}
\newcommand{\cfbc}{{\cal F}_{B \leftrightarrow C}}
\newcommand{\cc}{{\cal C}}
\newcommand{\ccbc}{{\cal C}_{B \leftrightarrow C}}

\newcommand{\lr}{\leftrightarrow}
\newcommand{\rl}{\leftrightarrow}
\newcommand{\pbx}{\bp \rightarrow -\bp}
\newcommand{\sqq}{{S_{Q\bar{Q}}}}
\newcommand{\Eta}{{\Omega}}
\newcommand{\figone}{{$1$}}
\newcommand{\figtwo}{{2}}
\newcommand{\figthree}{{3}}
\newcommand{\ti}{{\otimes}}
\newcommand{\eh}{{\hspace{.7cm}}}
\newcommand{\ehh}{{\hspace{.3cm}}}
\input epsf

\begin{document}
\title{\begin{flushright} \Large 
\small{hep-ph/9611375} \\
\small{MC-TH-96/26} \end{flushright} 
\vspace{0.6cm}  
\LARGE \bf  Why Hybrid Meson Coupling to Two S--wave Mesons is Suppressed}
\author{Philip R. Page\thanks{\small \em E-mail: prp@jlab.org} \thanks{\small \em Present address:
Theory Group, Thomas Jefferson National Accelerator Facility, 
12000 Jefferson Avenue, Newport News, VA 23606, USA.}\\
{\small \em Department of Physics and Astronomy, University of Manchester,} \\
{\small \em Manchester M13 9PL, UK}}
\date{November 1996}
\maketitle
\abstract{
We introduce strong interaction selection
rules for the two--body decay and production of hybrid and
conventional mesons coupling to two S--wave hybrid or conventional mesons. The rules arise 
from symmetrization in states in the limit of non--relativistically moving quarks. 
The conditions under which the connected coupling of a hybrid to two
S--wave hybrid or conventional mesons is suppressed are determined
by the rules, and the nature of their breaking is indicated.}

\vspace{1cm}

Certain vanishing decays were first noted in model calculations, and speculated
to be of a common origin \cite{kalashnikova95}.
For example, for hybrids with a transverse 
electric constituent gluon (``TE hybrids'') \cite{penepsi}, 
and hybrids arizing in the limit where quarks move adiabatically
(slowly) with respect to the gluonic degrees of freedom (``adiabatic hybrids'') in the flux--tube 
model \cite{page95hadron}, vanishing decays to orbital angular
momentum $L = 0$ (S--wave) mesons were found. In both cases it was remarked
that this was true for non--relativistically moving quarks 
with the same S--wave meson spatial wave functions  
and where the quark and antiquark in the hybrid have identical constituent masses.
We show that the rules found in detailed dynamical models are included
in the novel, more general and model--independent statement of the rules 
provided by us. The exact conditions under which hybrid coupling
to S--wave mesons vanishes are uncovered. In order to demonstate
vanishing coupling, we develop two--body decay and production
selection rules for non--relativistically moving quarks. 
The rules will be found
to apply to hybrid and
conventional mesons coupling to two S--wave hybrid and conventional mesons, 
and arise due to symmetrization in the latter two states. 

We shall be interested in strong decay and production $A\rl
BC$ processes in the rest frame of A. For simplicity we usually
refer to the decay process $A\rightarrow BC$ (see Fig. 1), but the statements will be equally
valid for the production process $A\leftarrow BC$.
The states A, B and C contain a ``valence'' quark and antiquark and
arbitrary gluonic content, i.e. they are conventional or hybrid
mesons.
In this letter we restrict to S--wave states B and C,
which are either radially excited or ground states. Clearly conventional S--wave meson
ground states B and C, which are henceforth mentioned in the examples,
 are most likely 
to be allowed by phase space. We assume that states B and C are identical in all respects
except, in principle, their flavour, spin and their equal but
opposite momenta $\bp$ and $ -\bp$.
Hence B and C have the same radial and gluonic excitation,
as well as the same internal structure.

A given decay topology 
explicitly assumes a certain non--relativistic spin $\sqq$ at each vertex. The physical
applicability of the rules will hence be determined by the extent to which a given decay
topology is believed to dominate decays. Models of meson decay usually assume either pair creation
with spin $\sqq$ or the exchange of a single 
quantum, either scalar confining, or colour Coulomb one gluon
exchange (OGE) or transverse OGE (see Appendix B of ref. \cite{ackleh96}). 
These all, in the non--relativistic limit, involve $\sqq=1$ pair creation and
do not involve spin change for a forward moving quark\footnote{The only exception is
for a forward moving quark interacting with a transverse gluon,
where a spin 1 term contributes. The term is not considered in this letter.}. We shall analyse
the implications of
symmetrization for some of the model topologies in Figure 1,
especially
the connected topology 1. 

The non--relativistic symmetrization selection rule and the conditions
for vanishing hybrid couplings are clearly stated
in the next section, where we perform a detailed derivation. In
section 2 non--connected topologies are touched apon. Breaking of the
rules and phenomenology are discussed in section 3.

\begin{figure}
\begin{center}
\leavevmode
\hspace{-.3cm}\hbox{\epsfxsize=5 in}
\epsfbox{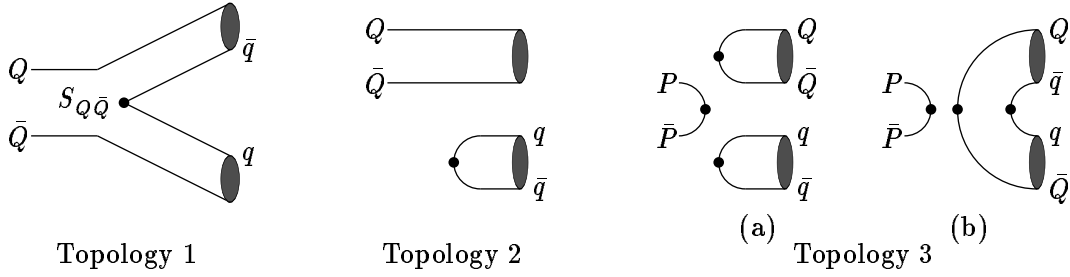}
\vspace{-.6cm}\end{center}
\caption{Hybrid and conventional meson decay topologies. For each diagram, state A is on the left--hand side, and
states B
and C on the right--hand side. Quark flavours are
labelled by $Q,\; q$ and $P$. The ``dots'' indicate vertices where 
pair creation with spin $\sqq$ is allowed.}
\end{figure}

\section{Connected coupling}

In this section we shall focus on topology \figone,
and shall follow the symmetrization arguments of ref. \cite{sel1}. 

For the strong processes we consider the helicity amplitude for the connected decay is always the
sum of two parts \cite{sel1}: 

\beqn
\cat (\bp)  = \cc\ti\cf\ti\cs\ti\cl (\bp)  +
\ccbc\ti\cfbc\ti\csbc\ti\clbc (\bp) 
\eeqn
Here we factored out the colour $\cc$, flavour $\cf$ and spin $\cs$
overlaps, and the ``remaining'' overlap $\cl$. $B
\lr C$ denotes the effect of formally exchanging labels that specify the states B and C.
When we exchange the momentum $\bp$  to $-\bp$, it is equivalent to exchanging the labels $B \lr C$ for every
property in the wave function of the state except flavour, colour and
spin, i.e. $\cl \rl  \clbc$  \cite{sel1}. 
Hence, under $\pbx$

\beqna
\lefteqn{\cat (\bp)  \rightarrow
\cc\ti\cf\ti\cs\ti\cl (-\bp)  + \ccbc\ti\cfbc\ti\csbc\ti\clbc (-\bp)
} \nonumber \\ & & 
= \cc\ti\cf\ti\cs\ti\clbc (\bp)  + \ccbc\ti\cfbc\ti\csbc\ti\cl (\bp) \nonumber  \\ & & 
= fs\; \{\ccbc\ti\cfbc\ti\csbc\ti\clbc(\bp) +
\cc\ti\cf\ti\cs\ti\cl(\bp)\} = fs\; \cat(\bp)
\eeqna 
We assumed that the colour structure of B and C are the same so that
$\ccbc = \cc$ \cite{sel1}. Also $\cfbc \equiv f \cf$ for the flavour scenarios we consider \cite{sel1}.
From the Appendix we see that $\csbc = s \cs$ where $s=(-1)^{S_A+S_B+S_C+\sqq}$. 
Here $S_A,\; S_B$ and $S_C$ are the spins of states A, B and C.
Until now we have referred to the helicity amplitude $\cat (\bp)$. Since its behaviour
under $\pbx$ is independent of total angular momentum projections, its behaviour
remains the same for a linear combination of amplitudes with various total angular momentum projections,
an example of which is a partial wave amplitude. By an abuse of notation we shall also call the
partial wave amplitude ``$\cat (\bp)$''. Hence $\pbx$ implies $\cat (\bp)\rightarrow fs\; \cat(\bp)$.
For decays where $fs = (-1)^{L+1}$, where $L$ is the partial wave
between B and C, we conclude that $\pbx$ implies
$\cat(\bp) \rightarrow (-1)^{L+1}\cat(\bp)$. 
Since in L--wave under $\pbx$ we have by analyticity that $\cat(\bp) \rightarrow (-1)^{L} \cat(\bp)$,
it follows that $\cat(\bp)$ vanishes. This is the desired result.

Now we shall find necessary and sufficient conditions for the
requirement $fs = (-1)^{L+1}$.
Since the B and C are identical (except for flavour
and spin) they have the same parity, due to the fact that the parity
of a state is fully determined by the intrinsic parities of the
constituents and the orbital excitation between them, {\it not} by the
spin of a state.
We conclude that for a parity allowed process, the parity of state A
is $P_{A} = (-1)^{L}$.  We shall see below that $fs = (-1)^{S_A+\sqq}
C_{A}^0$. For a neutral state,
$C_{A}^{0}$ is just the C--parity of the state. 
For charged states (with no C--parity), we assume that at least one of
the states in the isomultiplet
it belongs to has a well--defined C--parity, denoted by $C_{A}^{0}$.
We now have 

\beqn
P_{A}= (-1)^{L} = - (-1)^{L+1} = -fs = (-1)^{S_A+\sqq+1}C_{A}^0
\eeqn

Defining the action of charge conjugation on a spin 0 state $Q\bar{q}$ as 
$C(Q\bar{q}) = c\; q\bar{Q}$,
where $c$ is an intrinsic charge conjugation, and noting that
$C((Q\bar{q})^*) =  - c\; (q\bar{Q})^*$ for a spin 1 state $(q\bar{Q})^*$, we now show for various flavour scenarios that 
$fs = (-1)^{S_A+\sqq} C_{A}^0$.

{\bf Category I:} For a connected decay of the type $Q\bar{Q} \rightarrow Q\bar{q}\; 
q\bar{Q}, \; Q\bar{q} \; (q\bar{Q})^*, \; (Q\bar{q})^* \; (q\bar{Q})^*$ $\; (Q\neq q$), 
states B and C are not in general eigenfunctions of charge conjugation.
However, the linear combinations $Q_{C} \equiv
\frac{1}{\sqrt{2}}(|Q\bar{q}\rangle \; C \;  |q\bar{Q}\rangle)$ (spin 0) and $Q_{C}^* \equiv
\frac{1}{\sqrt{2}}(|(Q\bar{q})^*\rangle \; C \;  |(q\bar{Q})^*\rangle)$ (spin 1) have charge
conjugation $cC$ and $-cC$ respectively.  
In addition, in $Q\bar{Q}\rightarrow Q\bar{q}  \;
q\bar{Q}$ there is no flavour symmetry on exchange $B \lr C$, since one of the diagrams vanishes.
As we shall see, this
is corrected by considering decays into the eigenstates of charge conjugation $Q_{C}$ and $Q_{C}^*$.
We can decompose
$Q\bar{q} \; (q\bar{Q})^* = \frac{1}{2}(Q_+Q_+^* - Q_-Q_-^* - Q_+Q_-^*
+Q_-Q_+^*)$ into parts with proper $C$ symmetry.
The same can be done for $Q\bar{q}\; q\bar{Q}$ and $(Q\bar{q})^*\; (q\bar{Q})^*$.

Noting that $Q_{C}$ and $Q_{C}^*$ differ only in spin,
their flavour behaviour is identical, so we only discuss the flavour
behaviour of $Q_{C}$. 
$Q\bar{Q}\rightarrow Q_+Q_+,\; Q_-Q_-$ are trivially invariant under $B\rl C$, so that $f=1$. 
Also, for

\beqn
\plabel{fl}
Q\bar{Q} \rightarrow Q_+Q_- : \;\; \langle Q\bar{Q} | Q_+Q_- \rangle = -\frac{1}{2} \hspace{1.3cm}
Q\bar{Q} \rightarrow Q_-Q_+ : \;\; \langle Q\bar{Q} | Q_-Q_+ \rangle = \frac{1}{2}
\eeqn
and hence  $f=-1$ for $Q\bar{Q}\rightarrow Q_+Q_-,\; Q_-Q_+$. 

For $Q_+Q_+,\; Q_-Q_-,\; Q_+^*Q_+^*,\; Q_-^*Q_-^*,\; Q_+Q_-^*,\; Q_-Q_+^*$ states B and C we  
have $C_{B+C}=1$ and by conservation of charge conjugation $C_{A} = 1$. For 
the first four combinations $f=1$ so that $fs
=+(-1)^{S_A+\sqq} = (-1)^{S_A+\sqq} C_A$.
For the latter two combinations $f=-1$ and hence $fs
=-(-1)^{S_A+1+\sqq} = (-1)^{S_A+\sqq} C_A$.

For $Q_+Q_-,\; Q_-Q_+,\; Q_+^*Q_-^*,\; Q_-^*Q_+^*,\; Q_+Q_+^*,\; Q_-Q_-^*$ states B and C  
we have $C_{B+C}=-1$ which implies that $C_{A} = -1$. For the first four combinations 
$f=-1$ so that $fs=-(-1)^{S_A+\sqq} = (-1)^{S_A+\sqq} C_A$.
For the latter two combinations $f=1$ and hence $fs
=+(-1)^{S_A+1+\sqq} = (-1)^{S_A+\sqq} C_A$.

{\sf Examples:} $s\bar{s} \rightarrow K\bar{K},\; K\bar{K}^*,\;  K^*\bar{K}^*$;\eh $c\bar{c} \rightarrow
D\bar{D},\; D\bar{D}^*,\; D^*\bar{D}^*,\; D_s\bar{D}_s,\; D_s\bar{D}_s^*,\; D^*_s\bar{D}^*_s$;\eh  
$b\bar{b} \rightarrow B\bar{B},\; B\bar{B}^*,\; B^*\bar{B}^*,\;
B_s\bar{B}_s\; , B_s \bar{B}^*_s,\; B^*_s\bar{B}^*_s$;\eh 
$\Pi^{0} \rightarrow \pi^{+}\pi^{-},\; \pi^{\pm}\rho^{\mp},\; \rho^{+}\rho^{-}$;\eh \\
$\Eta \rightarrow \pi^{+}\pi^{-},\; \pi^{\pm}\rho^{\mp},\; \rho^{+}\rho^{-}$,

\noindent where e.g. $K\bar{K}$ includes $K^+K^-$ and $K^0\bar{K}^0$, 
$\Pi^{0}$ denotes a state with flavour $u\bar{u} - d\bar{d}$ and
$\Eta$ denotes $u\bar{u}+d\bar{d}$. 

{\bf Category II:} If we assume isospin symmetry for $u,d$ quarks, then for
G--parity eigenstates $G_{A} = G_{B}G_{C}$ by G--parity conservation. Since
$G_{H} = C_H^0 (-1)^{I_H}$, we obtain $C_{A}^0C_{B}^0C_{B}^0 = (-1)^{I_A+I_B+I_C}$. It can, however, be verified that $f=(-1)^{I_A+I_B+I_C}$ for
decays involving $u,d$ quarks (see the Appendix of ref. \cite{sel1}). This
continues to hold \cite{sel1} for $Q\bar{Q}\rightarrow Q\bar{Q}\; Q\bar{Q}$ with $Q = s,c,b$
(where $f=1$).

For $Q\bar{Q}\; Q\bar{Q},\; (Q\bar{Q})^*\; (Q\bar{Q})^*$ states B and C we have
$C_B^0C_C^0 = 1$ so that $C_{A}^0 = (-1)^{I_A+I_B+I_C}$ and hence 
$fs = (-1)^{I_A+I_B+I_C}(-1)^{S_A+S_B+S_C+\sqq} = 
(-1)^{S_A+\sqq}C_A^0$.

For $Q\bar{Q}\; (Q\bar{Q})^*$ states B and C we have
$C_B^0C_C^0 = -1$ so that $C_{A}^0 = (-1)^{I_A+I_B+I_C+1}$ and hence 
$fs = (-1)^{I_A+I_B+I_C}(-1)^{S_A+S_B+S_C+\sqq} = (-1)^{S_A+\sqq}C_A^0$.

{\sf Examples:}   $\Eta,\;  s\bar{s} \rightarrow \eta\eta,\; \eta^{'}\eta,\; \eta\omega, \;  \eta^{'}\omega,\; 
\eta\phi,\; \eta^{'}\phi,\; \pi^{0}\pi^{0},\; \pi^{0}\rho^{0},\;
 \rho^{0}\rho^{0}, \; \omega\omega,\;  \phi\phi$; \eh \\ $\Pi^{0} \rightarrow \pi^{0} \eta,\; 
\pi^{0} \eta^{'},\;  \pi^{0} \omega,\;  \rho^{0} \eta,\;  \rho^{0}
\eta^{'}, \; \rho^{0} \omega$; \eh \\
$\Pi^{\pm} \rightarrow \pi^{\pm}\pi^{0},\;  \pi^{\pm}\eta,\;   \pi^{\pm}\eta^{'},\;  \pi^{\pm}\omega,\; 
\pi^{\pm}\rho^{0}, \; \rho^{\pm}\pi^{0},\;  \rho^{\pm}\rho^{0},\;  \rho^{\pm}\eta,\;
 \rho^{\pm}\eta^{'},\; \rho^{\pm}\omega$.

\vspace{0.13cm}

{\bf Non--relativistic Symmetrization Selection Rule:} {\it Connected
decay and production of $C^0_AP_A =$ $(-1)^{S_A+\sqq+1}$
hybrid and conventional mesons with quark content $Q\bar{Q}$, 
where $S_{A}$ is the spin of the state and $\sqq$ the spin of
the created or annihilated pair, 
coupling to S--wave hybrid or conventional mesons which are identical
in all respects, expect possibly flavour and spin,
vanish for non--relativistic quarks. In the case of isospin symmetry
for topologies involving only $u,d$ quarks 
the same rule applies in addition to states A with quark content $Q\bar{q}$.}

\vspace{0.13cm}

If $\sqq = 1$ as models suggest, the condition that $C_{A}^0P_{A}=(-1)^{S_A+\sqq+1}$ 
is fulfilled for $J_A^{P_AC_A^0} = (0,1,2,\ldots)^{\pm\pm}$ 
with $S_A = 0$ and $J_A^{P_AC_A^0} = (0,1,2,\ldots)^{\pm\mp}$ with $S_A = 1$.

For conventional mesons, $C_{A}^0 = (-1)^{L_A+S_A}$ and $P_A =
(-1)^{L_A+1}$, so that $C_{A}^0P_A = (-1)^{S_A+1}$. So only spin 0
pair creation, usually thought to be highly suppressed, gives vanishing amplitudes. 
This may be related to the phenomenological success \cite{geiger94} of decay models fitting
experiment with spin 1 pair creation, 
although theoretical considerations independently motivate spin 1 pair creation. $J^P = 1^-$ transverse magnetic constituent gluons yield 
$1^{+-},\; (0,1,2)^{++}$ hybrids (``TM hybrids''), which have identical spin structure to
their conventional orbitally excited meson $J^{PC}$ partners, so their spin 0 decays vanish, but
not their spin 1 decays, consistent with covariant oscillator quark model calculations which 
find non--zero TM hybrid decays to S--wave mesons for spin 1 pair creation \cite{ishida89}.
Moreover, ref. \cite{penesize} finds $1^{-+} \rightarrow \rho\pi = 300$ MeV, 
with $S_A = 0$ quarks and spin 1 pair creation in the non--relativistic approximation.

If we assume spin 1 pair creation to be
dominant, $P_{A} = (-1)^{S_A+\sqq+1} C_{A}^0$ is satisfied for 
adiabatic hybrids\footnote{The listed $J^{PC}$ are those of the
lowest lying hybrids in the $E_u$ flux--tube representation on the hypercubic lattice
\protect\cite{perantonis90}. For spin 1 pair creation, $P_{A} = (-1)^{S_A+\sqq+1} C_{A}^0$ is also satisfied for energetically higher lying hybrids in the $E_u, A_{1u}, A_{2u}, B_{1u}$ and $B_{2u}$ representations. This is independent of the orbital and radial excitation of the
$Q\bar{Q}$, and fully determined by the flux--tube representation, as can most
easily be seen in the flux--tube model, where $P_{A} = (-1)^{S_A+\sqq+1} C_{A}^0$
is equivalent to a constraint on the flux--tube degrees of freedom 
\protect\cite[Eqns. A1 - A2]{paton85}.} 
with ${J_A}^{P_AC_A^0}_{\;\: S_A} = (0,1,2)^{-+}_{1}, 1^{--}_{0}, (0,1,2)^{+-}_{1},
1^{++}_{0}$ and for the $J^P = 1^+$ TE hybrids $(0,1,2)^{-+}_{1},
1^{--}_{0}$.  Note that the latter hybrids have the same
spin structure as their adiabatic limit $J^{PC}$ partners. Hence,

\vspace{0.13cm}

\noindent {\it Connected decay and production of adiabatic hybrids coupling to S--wave conventional mesons which are
identical in all respects, 
expect possibly flavour and spin,
vanish for non--relativistic quarks with spin 1 pair creation. The
quark content of the hybrid is either $Q\bar{Q}$ or $Q\bar{q}$, where the
latter is only relevant to topologies involving only $u,d$ quarks where
isospin symmetry is assumed. } 

\vspace{0.13cm}

\noindent Vanishing decay of adiabatic hybrids to S--wave mesons via spin 1 pair creation is confirmed in the flux--tube model \cite{page95hadron,page95light}
and for TE hybrids\footnote{Ref. \protect\cite{tanimoto83} claims in a specific model that the results survive even 
after the lifting of the assumption of non--relativistic quark motion.} in refs. 
\cite{penepsi,gutsche92,tanimoto82,tanimoto83}.
It was historically surprising that vanishing decays occur
in models motivated by both the strong and weak
coupling limits of QCD.
We have shown that this is because the decays have the same spin structure.
TE and adiabatic hybrids can have connected decay via spin 0 pair creation,
although in the
case of $0^{+-},1^{-+},2^{+-}\rightarrow$ identical $J=0$ states we know \cite{sel1} 
that the decays vanish by the symmetrization rules of ref. \cite{sel1}.
TE and adiabatic hybrids with non--exotic $J^{PC}$ have the opposite spin $S_A$ to their
conventional meson partners. It is hence clear why vanishing decays with
$P_{A} = (-1)^{S_A+\sqq+1} C_{A}^0$ arise {\it either} for hybrids {\it or} mesons, 
depending on $\sqq$.

\section{Non--connected coupling}

Except for the decays $\Pi^{\pm} \rightarrow \pi^{\pm}\pi^0,\; \rho^{\pm}\pi^0, \; 
\pi^{\pm}\rho^0,\; \rho^{\pm}\rho^0$ 
all decays listed in the previous section have contributions from non--connected
topologies 2 and 3.

Category II non--relativistic symmetrization arguments can be applied to the non--connected 
topology \figtwo\, yielding no new vanishing decays not included amongst those listed
for the same flavour topology in ref. \cite{sel1}.

Symmetrization arguments can be applied for topologies \figthree\ 
to yield vanishing decays only in specific models. In other models, the diagrams with $B\rl C$ 
are not topologically distinct, making it impossible to proceed.
No phenomenologically successful models have to the best of our knowledge been proposed utilizing 
topologies \figthree. In the light of this, we do not
develop symmetrization arguments further.


\section{Phenomenology}

There are considerably more sources of breaking of the non--relativistic
symmetrization selection rules than for those discussed in ref. \cite{sel1}.
As we shall see, breaking of
selection rules is likely to be the smallest for decays of $c\bar{c}$ and $b\bar{b}$ states.
Breaking of rules arise due to:

$\bullet$ Differing on--shell wave functions for B and C: 
Final states with different internal structure would break the symmetrization selection rules.
There is, however, no explicit breaking due to differences in energy or mass \cite{sel1}.
Corrections due to differing spatial wave functions for B and C due to
spin--dependent forces
are found to be given by $(R_B^2-R_C^2)^2 / (R_B^2+R_C^2)^2$ 
in models with harmonic oscillator wave functions \cite{penesize,page95light,kalashnikova94}, where
$R$ is the radius of the state. This ratio
ranges from approximately $20\%$ for $\rho\pi$ \cite{page95light} to $4\%$ for
$D^*\bar{D}$ or $1\%$ for $B^*\bar{B}$ to $0\%$ for $D\bar{D}, D^*\bar{D}^*, B\bar{B}, B^*\bar{B}^*$ \cite{page95thes}.
Topology 1 decay widths for $c\bar{c}$ and $b\bar{b}$ adiabatic hybrids to S--wave mesons of respectively $1-10$ MeV and $1-4$ MeV
have been predicted for spin 1 pair creation \cite{page95thes}.

$\bullet$ Differing off--shell wave functions for B and C: Breaking of the rules could
be more substantial \cite{sel1,page95light} than for on--shell states,
enabling off--shell meson exchange as a 
potentially significant hybrid
production mechanism, e.g in $\pi N\rightarrow 1^{-+}N$ with $\rho$ exchange.

$\bullet$ Different spin assignments and mixing: The spin assignments
of unmixed hybrid and conventional mesons assume that the
adiabatic limit survives even for light quark mesons. This is motivated
by the success of the non--relativistic quark model. Also, a simulation in the
flux--tube model indicates that
``mixing between [adiabatic] surfaces is of the order of 1\% or less'' 
even for light quark systems
\cite{merlin85}. Moreover, constituent gluon models usually find small mixing
between hybrid and conventional mesons
\cite{penepsi,tanimoto83}. 

$\bullet$ Relativistic effects: Nonwithstanding the successes of the non--relativistic quark model,
there is no decay of current interest for which the 
non--relativistic assumption is compelling. Even for a 
$b\bar{b}$ state decaying via s--quark pair creation, the s--quark is not manifestly
non--relativistic. If fully relativistic QCD sum rule calculations are a guide, $u,d$ quark 
$1^{-+}$ has a width of  $10 - 600$ MeV to $\rho\pi$ and $8 - 300$ MeV to $K^*K$ \cite{sumrules}.  
These large uncertainties in widths unfortunately leave the size of relativistic effects unresolved.

Decays suppressed by selection rules in a given topology can receive other contributions from:

$\diamond$ Other quark topologies: Of most interest here is the additional contribution
of non--connected topologies to a suppressed connected topology. 
If non--connected
decays of hybrids are similar to that of mesons, 
they are expected to be suppressed, less than ${\cal O}\; (10^{-1}$ MeV) for $c\bar{c}$ 
and ${\cal O}\; (10^{-2}$ MeV) for $b\bar{b}$ states.
If QCD sum rule calculations are a guide, $u,d$ quark hybrid 
$1^{-+}$ has a $\eta\pi$ width of $0.3$ MeV and a $\eta^{'}\pi$ width of $3-4$ MeV \cite{sumrules},
suggesting the suppression of non--connected hybrid decays.

$\diamond$ Different decay processes to those in Figure 1 and pair creation
with $S_{Q\bar{Q}}\neq 1$, as discussed before.

It seems reasonable to assume that decays of hybrid
$c\bar{c}$ to $D\bar{D},D^*\bar{D},D^*\bar{D}^*,D_s\bar{D}_s,D^*_s\bar{D}_s,D^*_s\bar{D}^*_s$
and hybrid $b\bar{b}$ to $B\bar{B},B^*\bar{B},B^*\bar{B}^*,B_s\bar{B}_s,B^*_s\bar{B}_s,B^*_s\bar{B}^*_s$
are suppressed. Moreover, 
hybrid $0^{+-},1^{-+},$ $2^{+-},3^{-+},\ldots$\ehh $c\bar{c}\rightarrow D\bar{D},D_s\bar{D}_s$ and 
$b\bar{b}\rightarrow B\bar{B},B_s\bar{B}_s$ vanish by $CP$ conservation. Since heavy hybrid states have consistently been 
predicted below the $(L=0)+(L=1)$ threshold \cite{perantonis90}, this implies the tantalizing prospect
of narrow $c\bar{c}$ and $b\bar{b}$ hybrids.

The principle of symmetrization applied to non--relativistic quarks has been shown to 
underpin the suppression of hybrid coupling to S--wave states for spin
1 pair creation, binding together numerous model calculations. 

Discussions with M.C. Birse, F.E. Close, N. Isgur, C. Michael, O. P\`{e}ne,
J.C. Raynal, M.R. Pennington, P. Sutton and motivation from S.-F. Tuan are acknowledged.

\appendix

\section{Appendix: Spin Overlaps}

The spin state is 

\beqn
|H\rangle = \sum_{h\bar{h}} H_{h\bar{h}} |h\rangle |\bar{h}\rangle
\hspace{1cm} \mbox{where}\hspace{.3cm}  H_{h\bar{h}} = \langle S_H S_H^z | \frac{1}{2}h \frac{1}{2}-\bar{h}\rangle
(-1)^{\frac{1}{2}-\bar{h}}
\eeqn
and $|\frac{1}{2}\rangle = \;\uparrow,\; |-\frac{1}{2}\rangle = \;\downarrow,\; 
|\bar{\frac{1}{2}}\rangle= \bar{\downarrow}$ and 
$|-\bar{\frac{1}{2}}\rangle = -\bar{\uparrow}$. This just yields the usual $S_H=1$ spin
$\uparrow\bar{\uparrow}, \;\frac{1}{\sqrt{2}}(\uparrow\bar{\downarrow}+\downarrow\bar{\uparrow}), \; \downarrow\bar{\downarrow}$ for $S_H^z = 1,0,-1$ and $\frac{1}{\sqrt{2}}(\uparrow\bar{\downarrow}-\downarrow\bar{\uparrow})$ for $S_H=0$. 
The pair creation or annihilation operator is 
$\sum_{h\bar{h}}\langle S_{Q\bar{Q}} S_{Q\bar{Q}}^z 
| \frac{1}{2}h \frac{1}{2}-\bar{h}\rangle(-1)^{\frac{1}{2}-\bar{h}}
|h\rangle |\bar{h}\rangle$. The advantage of this representation for spin 
is that a spin 0 operator is $\frac{1}{\sqrt{2}}\delta_{h\bar{h}}$ which is a multiple of
the identity matrix. So spin 0 interactions on forward moving quark lines are trivial.
The spin 1 operators are just related to the Pauli matrices as $-\frac{1}{2}(\sigma_x+i \sigma_y)$ for
$S_{Q\bar{Q}}^z = 1$, $\frac{1}{\sqrt{2}}\sigma_z$ for $S_{Q\bar{Q}}^z = 0$ and
$\frac{1}{2}(\sigma_x-i \sigma_y)$ for $S_{Q\bar{Q}}^z = -1$.

For the connected decay of topology \figone\ the spin overlap $\cs$ is 

\beqna
\lefteqn{
\sum_{a\bar{a}b} A_{a\bar{a}}B_{a\bar{b}} 
\langle S_{Q\bar{Q}} S_{Q\bar{Q}}^z | \frac{1}{2}b \frac{1}{2}-\bar{b}\rangle (-1)^{\frac{1}{2}-\bar{b}}
C_{b\bar{a}}    \nonumber } \\ & & 
= \sum_{a\bar{a}b\bar{b}} \langle S_A S_A^z | \frac{1}{2}a \frac{1}{2}-\bar{a}\rangle
\;\langle S_B S_B^z | \frac{1}{2}a \frac{1}{2}-\bar{b}\rangle
\;\langle S_{Q\bar{Q}} S_{Q\bar{Q}}^z | \frac{1}{2}b \frac{1}{2}-\bar{b}\rangle
\;\langle S_C S_C^z | \frac{1}{2}b \frac{1}{2}-\bar{a}\rangle
\eeqna 
which can easily be shown to give the sign
$s=(-1)^{S_A+S_B+S_C+S_{Q\bar{Q}}}$ under $B\rl C$.

\end{document}